\documentclass[showpacs,preprintnumbers,amsmath,amssymb,showpacs]{revtex4}

\usepackage{graphicx}
\usepackage{dcolumn}
\usepackage{bm}
\usepackage{amssymb}
\usepackage{color}

\def\bea{\begin{eqnarray}}
\def\ena{\end{eqnarray}}

\begin{document}

\title{Limits on High-Frequency Gravitational Wave Background from its interplay with Large Scale Magnetic Fields}


\author{M.~S.~Pshirkov}
\affiliation{Pushchino Radio Astronomy Observatory, Astro Space Center, Lebedev Physical Institute, Pushchino, Russia\footnote[1]{e-mail: Pshirkov@prao.ru}}
\author{D.~Baskaran}
\affiliation{School of Physics and Astronomy, Cardiff University, Cardiff CF24 3AA, UK\\
and Wales  Institute of Mathematical and Computational Sciences, Swansea SA2 8PP United Kingdom \footnote[2]{e-mail: Deepak.Baskaran@astro.cf.ac.uk}}


\small


\begin{abstract}
In this work, we analyze the implications of graviton to photon conversion in the presence of large scale magnetic fields. We consider the magnetic fields associated with galaxy clusters, filaments in the large scale structure, as well as primordial magnetic fields. {We analyze the interaction of these magnetic fields with an exogenous high-frequency gravitational wave (HFGW) background which may exist in the Universe. We show that, in the presence of the magnetic fields, a sufficiently strong HFGW background would lead to an observable signature in the frequency spectrum of the Cosmic Microwave Background (CMB).} The sensitivity of current day CMB experiments allows to place significant constraints on the strength of HFGW background, $\Omega_{GW}\lesssim1$. These limits are about 25 orders of magnitude stronger {than currently existing direct constraints} in this frequency region.
\end{abstract}



\today

\maketitle


\section{Introduction \label{SectionI}}

In recent times there has been a rising interest  in
high-frequency gravitational waves (HFGWs), i.e.~waves with
frequencies higher than $\nu \gtrsim10^5~{\rm Hz}$. Although most
astrophysical sources radiate gravitational waves at much lower
frequencies $\nu \lesssim 10^{3} ~{\rm Hz}$
\cite{Grishchuk2001,Cutler2002,Schutz2009}, the high frequencies
might contain gravitational wave signal coming from the very early
Universe as well as some other sources 
{and mechanisms such as cosmic strings, evaporation of light primordial 
black holes and effects associated with presence of higher dimensions}
\cite{Gasperini1993,Brustein1995,Gasperini2003,Bisnovatiy-KoganRudenko2004,Grishchuk2007,Giovannini2009,Clarkson2007,
Anantua2008,Easther2007}. Currently there is considerable interest
in the possibility of building HFGW detectors capable of detecting
these signals as well as signals created in laboratory
\cite{GrishchukSazhin1974,BraginskyRudenko1978,Grishchuk2003,
Baker2005,Cruise2006,Nishizawa2008,BakerWebsite}. In light of the
rising interest in HFGW it is instructive to analyze the possible
observational constraints on the HFGW background. Existing {direct}
observational constraints on HFGWs come from laser-inteferometer
type experiment and are not very restrictive,
$\Omega_{GW}\lesssim10^{26}$ at 100 MHz frequency \cite{Akutsu2008}. In
this paper, in order to place constraints on the HFGWs, we shall
consider their possible signature in the cosmic microwave
background (CMB) due to their interaction with large scale
magnetic fields in the Universe.

Gertsenshtein \cite{Gertsenshtein1962} {(see also \cite{Braginskii1974,Grishchuk1980,Zeldovich1983})} showed that  in a
stationary electromagnetic field gravitons may decay into photons. 
A graviton propagating in a stationary electromagnetic field may interact with virtual photons of that field, and produce a real photon with almost the same frequency
and wavevector as the original graviton (see \cite{Fargion1996} for a modern exposition). In the framework of classical field theory
the graviton to photon conversion can be understood as a result of the interaction of a time varying metric perturbation field with 
a stationary electromagnetic field, leading to time variations in the latter, i.e. production of photons.
In this paper, we shall analyze the observational consequences of the possible decay of
gravitons into photons in the presence of magnetic fields with a
view to place constraints on the HFGW background. There is
currently ample evidence for widespread existence of magnetic
fields in the Universe {\cite{Kronberg1994,Beck1996,Grasso2001,Widrow2002}}.
The magnetic fields are known to exist in a wide variety of
scales. The galactic magnetic fields have a characteristic
strength of $\sim 1 ~\mu{\rm G}$ and coherence scales of a few
kiloparsecs (kpc). In clusters of galaxies, the magnetic fields
have a typical strength of $1-10~\mu{\rm G}$ and coherence lengths
of $10-100~{\rm kpc}$ \cite{Dar2005}. Of interest are the magnetic
fields with field strength $\sim 0.3~\mu{\rm G}$ and coherence
lengths of $\sim 1~{\rm Mpc}$ observed in the galaxy overdense
filaments of typical size $\sim 50~{\rm Mpc}$ in the large scale
structure (LSS) \cite{Xu2006}. Furthermore, there are strong
reasons to believe that at the largest scales there exist magnetic
fields of primordial origin \cite{Grasso2001}. The tightest
constraints on the strength of primordial magnetic fields (PMF)
come from the analysis of anisotropies in the CMB and are limited
to the present day by value of $\lesssim 10^{-9}-10^{-8}~{\rm G}$
{\cite{Grasso2001,Durrer2006,Giovannini2008,Kristiansen2008,Yamazaki2008,Kahniashvili2008}}.

The existence of these magnetic fields allows to place
observational constraints on the strength of the possible HFGW
background. In the presence of magnetic fields a sufficiently
strong HFGW background would lead to the production of photons
through the Gertsenshtein effect that could be observed as
distortions in the frequency spectrum of the CMB. On the other
hand, the absence of these distortions would signify an upper limit
on the strength of the HFGW background. In the present work we
shall estimate the magnitude of the expected spectral distortions
in the CMB and as a consequence analyze the possible constraints
on HFGWs. Before proceeding to the main topic of the current paper, it is worth pointing out that the large scale magnetic fields could themselves produce significant gravitational wave background \cite{Durrer2000,Caprini2006}. These gravitational waves would leave their imprint in the temperature and polarization anisotropies of the CMB primarily at large angular scales corresponding to multipoles $\ell\lesssim100$ \cite{Mack2002,Lewis2004,Subramanian2006}. However, in the present paper, we shall restrict our analysis to the interaction of an exogenous HFGW background with large scale magnetic fields.


\section{The probability of graviton to photon conversion\label{SectionII}}

In a uniform magnetic field characterized by strength $B$ the probability of conversion of a
graviton, travelling perpendicular to the magnetic field lines, into
a photon is given by \cite{Cillis1996}
\begin{eqnarray}
P_{g\rightarrow \gamma} \simeq 8.3\times
10^{-50}\left(\frac{B}{1{\rm G}}\right)^2\left(\frac{L_{\rm
coh}}{1{\rm cm}}\right)^2. \label{equation1}
\end{eqnarray}
In the above expression $L_{\rm coh}$ is the coherence length for the graviton to photon conversion process. In perfect vacuum, the coherence length $L_{\rm coh}$ is equal to the length of coherence of the magnetic field, i.e. distance over which the magnetic field remains homogenous. However, in the situations considered in the current work, the coherence length $L_{\rm coh}$ is determined primarily by the plasma effects. In presence of plasma, the velocity of photons differs from the graviton velocity. For this reason, the condition for resonant conversion of gravitons into photons will typically hold for shorter distances than in the case of pure vacuum (see Eqs.~(16,17) in \cite{Fargion1996}). The coherence length in the presence of plasma is given by the expression \cite{Fargion1996}
\begin{eqnarray}
L_{\rm coh} \simeq 3\times10^{14}\left(\frac{f}{10^{10}{\rm Hz}}\right)\left(\frac{n_e}{1{\rm cm}^{-3}}\right)^{-1} {\rm cm},
\label{equation2}
\end{eqnarray}
where $n_e$ is the electron density and $f$ is the frequency of the graviton as well as the subsequently created photon. {In the above expression and through out the paper we use $10^{10}~\rm{Hz}$ as the referential frequency since it corresponds to the theoretically predicted high frequency end of the spectrum of relic gravitons.} 

In general, the coherence length $L_{\rm coh}$ is significantly smaller than the total linear dimensions of the magnetic field structure $L_{\Sigma}$. The total number of coherent domains is given by the ratio $\eta = L_{\Sigma}/L_{\rm coh}$. Hence, the total probability of graviton-to-photon conversion in the magnetic field structure of length $L_\Sigma$ is given by
\begin{eqnarray}
\mathcal{P}_{g\rightarrow\gamma} &\simeq& \eta P_{g\rightarrow \gamma}, \nonumber \\
&=& 7.2\times 10^{-11}\left(\frac{B}{1{\rm G}}\right)^2\left(\frac{f}{10^{10}{\rm Hz}}\right)\left(\frac{n_e}{1{\rm cm}^{-3}}\right)^{-1} \left(\frac{L_{\Sigma}}{1{\rm Mpc}}\right).
\label{equation3}
\end{eqnarray}

\subsection{Magnetic fields in galaxy clusters and filaments}

Let us analyze the conversion probabilities for magnetic fields
associated with galaxy clusters and the magnetic fields
in filaments. For estimating the probability of
graviton-to-photon conversion in magnetic fields associated with
galaxy clusters, we shall take the typical value
$L_{\Sigma}=2~{\rm Mpc}$, $n_e=10^{-5}~{\rm cm}^{-3}$ and
$B=3~\mu{\rm G}$ for the characteristic size of the galaxy
cluster, its mean electron density and its characteristic magnetic
field strength, respectively \cite{Dar2005}. Substituting these
values into (\ref{equation3}) we get
\begin{eqnarray}
\mathcal{P}_{g\rightarrow\gamma} \left({\rm galaxy~cluster}\right)
\simeq 1.4\times10^{-16} \left(\frac{f}{10^{10}{\rm Hz}}\right) n_{\rm GC},
\label{Pgalaxycluster}
\end{eqnarray}
where $n_{\rm GC}$ is the number  of galaxy clusters along the
line of sight. In the case of filaments, we set
$L_{\Sigma}=50~{\rm Mpc}$, $n_e=10^{-7}~{\rm cm}^{-3}$ and $B=0.3~\mu{\rm G}$ correspondingly. In this
case we arrive at a somewhat larger probability
\begin{eqnarray}
\mathcal{P}_{g\rightarrow\gamma} \left({\rm filament}\right)
\simeq 3.2\times10^{-15}  \left(\frac{f}{10^{10}{\rm Hz}}\right) n_{\rm F} ,
\label{Pfilament}
\end{eqnarray}
where $n_{\rm F}$ is the number of filaments along the line of sight. Simple estimations
\footnote{Numbers of filaments crossed is roughly equal to ratio of total path length inside filaments to the characteristic size of a single filament. The total path length inside filaments is equal to total distance from last scattering surface multiplied by filaments ``concentration", $d_f\simeq 4~ \mathrm{Gpc}\times\left(\frac{L_f}{L_v+L_f}\right)^3$, where $L_f$ is the characteristic size of filament and $L_v$ is the typical size of voids. Setting $L_f = 50~\mathrm{Mpc}$ and $L_v=100~\mathrm{Mpc}$ we arrive at $d_f\simeq 150~{\rm Mpc}$, and consequently $n_f\simeq 3$.}
suggest that the factor $n_{\rm F}$ could reach values $\sim 3-5$. However, to avoid speculations, in our estimation below we shall set $n_{\rm GC}=n_{\rm F}=1$. It is worth mentioning that in estimation of (\ref{Pgalaxycluster}) and (\ref{Pfilament}) we have assumed that the magnetic field is always pointing orthogonal to the line of sight. It is reasonable to assume that an exact evaluation involving appropriate averaging over the direction of the magnetic field would lead to a smaller probability but would not qualitatively change the result.

\subsection{Primordial magnetic fields}

Let us estimate the graviton-to-photon conversion probability for primordial magnetic fields. In estimating the probability in the case of PMF the cosmological expansion and the associated decay of these magnetic fields must be taken into account. With the expansion of the Universe the magnetic field scales in the following manner
\begin{eqnarray}
B(z) \simeq B_0(1+z)^2,
\nonumber
\end{eqnarray}
where $B_0$ is the characteristic value of primordial magnetic field at the
present epoch, and $z$
is the cosmological redshift. The coherence length scales
correspondingly as
\begin{eqnarray}
L_{\rm coh}(z) &\simeq& L_{\rm coh}(z_{rec})\left(\frac{1+z_{rec}}{1+z}\right)^{2}
\nonumber \\
&\simeq& 3.9\times10^{18} \left(\frac{1+z_{rec}}{1+z}\right)^{2} \left(\frac{f}{10^{10}{\rm Hz}}\right)~{\rm cm}.
\nonumber
\end{eqnarray}
In the above expression the coherence scale length just
after the epoch of recombination $L_{\rm coh}(z_{rec})$ was
calculated from (\ref{equation2}) setting $n_e(z_{rec}) =
x_{ion}\rho_{crit}\Omega_B(1+z_{rec})^3/m_p$, assuming a residual
ionization fraction $x=3\times10^{-4}$ \cite{Peebles1993}, and
setting $\Omega_B=0.04$, $\rho_{crit}=1.1\times10^{-29}{\rm
gm\cdot cm^{-3}}$. Note that, in the above expression and elsewhere
in the text, $f$ represents the frequency of gravitons/photons at
the present epoch. From the above expression it follows that the
conversion probability in a single coherence domain
(\ref{equation1}) is independent of the redshift
\begin{eqnarray}
P_{g\rightarrow \gamma} = 1.3\times10^{-18} \left(\frac{B_0}{10^{-9}{\rm G}}\right)^2 \left(\frac{f}{10^{10} {\rm Hz}}\right).
\nonumber
\end{eqnarray}
Thus, in order to estimate the total probability we need
to calculate the total number of coherence domains crossed by a
graviton. A graviton propagates through a single coherence scale
in a time period $\Delta t(z) \simeq L_{\rm coh}(z)/c$. Assuming a
matter dominated cosmological evolution, i.e. $1+z \simeq
\left(\frac{3}{2}H_0t\right)^{-2/3}$ where $H_0$ is the present
day Hubble constant, we arrive at the following integral for total
number of coherent domains
\begin{eqnarray}
\eta &=& \frac{c}{H_0L_{\rm coh}(z_{rec})\left(1+z_{rec}\right)^2}\int_{z_{min}}^{z_{max}}\frac{dz}{\sqrt{1+z}} \nonumber \\
&\simeq& \frac{2c}{H_0L_{\rm coh}(z_{rec})}\frac{\sqrt{1+z_{max}}}{(1+z_{rec})^2}.
\nonumber
\end{eqnarray}
Since we are primarily interested in observational manifestations
of graviton to photon conversion in CMB, we shall set $z_{max} =
10^3$ and $z_{min} = 10$ corresponding to the redshift of
recombination and reionization respectively. Since
the Universe was optically thick to CMB radiation prior to
recombination, the signature of any graviton to photon conversion
from an earlier epoch would not be seen. On the other hand, after
reionization the coherence length dramatically reduces due to the
increase in the density of free electrons $n_e$ (see
(\ref{equation2})), and the conversion probability becomes
negligible. Numerical evaluation leads to $\eta \simeq
2\times10^5$. Hence, the total probability of conversion is given
by
\begin{eqnarray}
\mathcal{P}_{g\rightarrow\gamma} \left({\rm primordial}\right) &=& \eta P_{g\rightarrow\gamma}
\nonumber \\
& \simeq & 2.5 \times 10^{-13}\left(\frac{B_0}{10^{-9}{\rm G}}\right)^2 \left(\frac{f}{10^{10} {\rm Hz}}\right).
\label{Pprimordial}
\end{eqnarray}
As can be seen, for a characteristic value of $B_0 = 10^{-9}~{\rm G}$ for the present day strength of the PMF, the conversion probability is almost two orders of magnitude larger than in the case of filaments.


\section{Observational implications\label{SectionIII}}

\subsection{Electromagnetic signal due to graviton to photon conversion}

Let us now estimate the expected electromagnetic signal
due to the considered graviton-to-photon conversion. The
electromagnetic energy flux $S_{EM}$ would be proportional to the
product of the gravitational wave energy flux $S_{GW}$ multiplied
by the total conversion probability
$\mathcal{P}_{g\rightarrow\gamma}$, i.e.~$S_{EM} \simeq
S_{GW}\mathcal{P}_{g\rightarrow\gamma}$. Assuming a statistically
isotropic gravitational wave background, the energy flux of the
gravitational wave field can be expressed in terms of its energy
density $S_{GW} = c\rho/4=c\Omega_{GW}\rho_{cr}/4$, where we have
introduced the gravitational wave fraction of the critical density $\Omega_{GW}$.
The expected electromagnetic flux is thus given by
\begin{eqnarray}
S_{EM} \simeq 7.2\times10^{-12}
\left(\frac{\mathcal{P}_{g\rightarrow\gamma}}{10^{-13}}\right)
\Omega_{GW} \frac{\rm erg}{{\rm cm}^2\cdot{\rm s}\cdot{\rm sr}}.
\nonumber\label{EMflux}
\end{eqnarray}
In order to compare the flux with the sensitivity of various experiments, it is convenient to express the result in terms of brightness temperature. The brightness temperature is related to the electromagnetic flux by the relation $\Delta T = c^2S_{EM}/2kf^3$. Thus, the expected electromagnetic signal is given by
\begin{eqnarray}
\Delta T \simeq 25
\left(\frac{\mathcal{P}_{g\rightarrow\gamma}}{10^{-13}}\right)\left(\frac{10^{10}\rm
{Hz}}{f}\right)^3\Omega_{GW} ~\mu{\rm K}. \label{DeltaTflux}
\end{eqnarray}
Comparing the flux for probabilities (\ref{Pgalaxycluster}), (\ref{Pfilament}) and (\ref{Pprimordial}), it can be seen that the strongest signal $\Delta T \sim 60\cdot \Omega_{GW}~\mu {\rm K}$ (assuming $B_0=10^{-9}~{\rm G}$ and $f= 10^{10}~{\rm Hz}$) is expected due to graviton conversion in the presence of PMF. Note that, the exact frequency dependence of the signal is determined by the frequency dependence of $\Omega_{GW}$. From (\ref{DeltaTflux}), (\ref{Pgalaxycluster}), (\ref{Pfilament}) and (\ref{Pprimordial}) it follows that, for a flat spectrum of HFGW (i.e.~$\Omega_{GW}\simeq {\rm const}$) the expected signal scales as $\Delta T \propto f^{-2}$ in terms of brightness temperature.

\subsection{Observational prospects and potential caveats}

In order to analyze the potential observational prospects,
it is instructive to compare the strength of the expected signal with the sensitivity of realistic detectors.
Recently, the AMI experiment \cite{AMI} achieved a sensitivity $\Delta T_{rms}
\simeq 1~{\mu}{\rm K}$ at a frequency $\nu\sim10^{10}~{\rm
Hz}$. In a typical Cosmic Microwave Background
experiment, at a frequency $\nu\sim10^{11}~{\rm Hz}$,
for a $\Delta\theta = 1^o$ resolution, the attainable sensitivity is $\Delta T_{rms}\simeq
1 ~\mu{\rm K}$ \cite{PlanckBlueBook}. The optimal frequency channel for constraining
HFGWs is a matter of trade-off between a signal weakening with
increase in frequency on the one hand, and a lower foreground
level at frequencies $\nu\sim10^{11}~{\rm Hz}$ (see for example
p.~4 in \cite{PlanckBlueBook}) on the other.
In our case, a sensitivity of $1~\mu\rm{K}$ at 10 GHz corresponds to
a sensitivity of $0.01~\mu\rm{K}$ at 100 GHz. Additionally, it is worth noting
that, potentially, the attainable sensitivity might be
considerably increased by increasing the time of observation. A CMB experiment typically
has to scan the whole sky, allowing for only $t_{pix}\sim10~\rm{sec}$ per individual pixel.
On the other hand if this time is increased to $t_{pix}\sim 1~{\rm
yr}$, the attainable sensitivity
would improve to $\Delta T_{rms}\simeq 5 \times 10^{-4} ~\mu{\rm
K}$ at $10^{11}~\rm{Hz}$. However, such an increase in observation time would require a
specially designed experiment dedicated solely to constraining
HFGW background.

\begin{figure}
\begin{center}
\includegraphics[width=10cm]{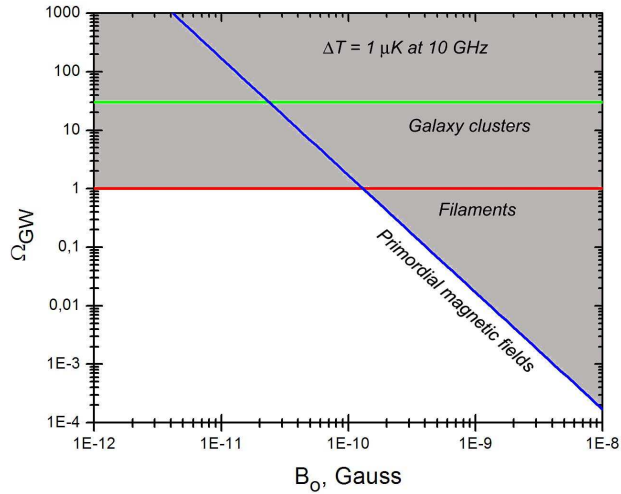}
\end{center}
\caption{The achievable constraints $\Omega_{GW}$ depending on the
strength of the primordial magnetic field $B_0$. For comparison
the horizontal lines represent the constraints due to magnetic
fields in galaxy clusters and filaments. The shaded area indicates
the region in parameter space that could be ruled out with current
observations. The sensitivity level is set to $1~\mu{\rm K}$ at $10~{\rm GHz}$
(equivalent to a sensitivity $0.01~\mu{\rm K}$ at $100~{\rm GHz}$),
and a red spatial spectrum for PMF is assumed.
}\label{figure1}
\end{figure}

Comparing the observational sensitivity with the
expected signal due to HFGWs in the CMB given by
(\ref{DeltaTflux}) in the context of PMF (\ref{Pprimordial}), in
the absence of a signal, we can place the following constraints on
HFGW background
\begin{eqnarray}
\Omega_{GW} \lesssim 1.7\times10^{-2} \left(\frac{\Delta
T_{rms}}{1~\mu{\rm K}}\right) \left(\frac{10^{-9}{\rm
G}}{B_0}\right)^2 \left(\frac{f}{10^{10}{\rm Hz}}\right)^2.
\label{HFGWconstraint}
\end{eqnarray}
On the other hand, HFGWs with $\Omega_{GW}$ larger than the threshold
value (\ref{HFGWconstraint}) would leave an observable signature
in the CMB. Note that, the constraints on $\Omega_{GW}$ crucially
depend on the strength of the PMF $B_0$. For a typical value
$B_0=10^{-9}~\rm{G}$ these constraints are 2-3 orders of magnitude
stronger than the analogous constraints due to magnetic fields in
galaxy clusters and filaments. In Figure \ref{figure1} we draw the
potential constraints on $\Omega_{GW}$ depending on the strength
of the PMF $B_0$. The shaded regions represent the regions in
$B_0$-$\Omega_{GW}$ space that could be potentially ruled out by
observations. For comparison, the two horizontal lines show
the constraints that arise when considering the magnetic fields in
galaxy clusters and filaments.

It is worth noting that in analyzing  the potential constraints on
$\Omega_{GW}$ through the process of graviton to photon conversion
in the presence of magnetic fields we have ignored the inverse
process of photon to graviton conversion. This inverse effect has
the same probability given by (\ref{equation1}). However, 
at frequencies $f\sim10^{10}~{\rm Hz}$, the energy density of CMB is several orders 
of magnitude smaller than the typical energy density of HFGW backgrounds considered in this work. 
For this reason, the total contribution of the inverse effect to changes in the electromagnetic flux remains subdominant.

A potential caveat in our  ability to constrain HFGWs arises due
to the differential nature of CMB measurements. The conversion
probability in presence of PMF is sufficiently isotropic, leading
to predominantly isotropic signal in $\Delta T$. The residual
anisotropic variations would be $\Delta T_{anis}\sim\Delta
T/\sqrt{\eta} = 3\times 10^{-3}\Delta T$. A conventional CMB
experiment would be restricted to ability to measure only this
residual anisotropic variations, weakening the potential
constraints on $\Omega_{GW}$. However, PMF produced during
inflation with a sufficiently red spatial spectrum
\cite{Gasperini1995}, may have significantly varying field
strength amplitudes in various domains of sub-horizon scale. For
these fields the conversion probability would be anisotropic
leading to a large anisotropy in the expected signal. On the other
hand, this isotropy problem would not arise when considering the
CMB signal due graviton conversion in magnetic fields in galaxy
clusters and filaments.

A further caveat is also worth mentioning here.
In order to detect or constrain the possible signal from HFGWs in
the CMB it is necessary to distinguish this signal from other
potential mechanisms contributing to the anisotropies in CMB. The
commonly considered contributions are the anisotropies due to
density perturbations and relic gravitational waves, anisotropies
due to Sunyaev-Zel'dovich (SZ) effect, and anisotropies arising
due to astrophysical foregrounds \cite{PlanckBlueBook}. However,
these contributions, in general, can be subtracted due to their
known frequency dependence. For example, it is known that, the
anisotropies due to density perturbations and relic gravitational
waves do not depend on frequency (in temperature units, in the Rayleigh-Jeans region). We can
estimate the SZ effect in filaments following \cite{Rephaeli1995}:
$\Delta T_{SZ} \simeq 2Ty \simeq 10^{-2}~\mu {\rm K}$ (where $y=
\int{dl~{\sigma_TkT_e}n_e/mc^2}$, and $T_e=10^6~{\rm K}$). This
signal has a well understood frequency dependence and for this
reason can also be subtracted. Finally, there are indications that the
various astrophysical foregrounds, that typically have an
amplitude $\Delta T_{foregrd}\sim 10^2\mu K$ at $\nu=10^{10}~{\rm
Hz}$, could be effectively subtracted to a level $\Delta T\lesssim1~\mu \rm{K}$
outside the galactic plane \cite{Hinshaw2007}.

{
Finally, it is useful to compare the sensitivity of the CMB experiments with other methods. 
The only existing direct measurements of the HFGW background, using laser-interferometeric 
type detectors, place an upper limit $\Omega_{GW}\lesssim10^{26}$ in the frequency range around 100~MHz \cite{Akutsu2008}.
Therefore, it seems highly unlikely that direct measurements would be able to compete with the sensitivity of CMB experiments in the foreseeable future.
The most stringent constraint on the possible strength of the HFGW background of cosmological origin 
are placed by the concordance with the Big-Bang Nucleosynthesis (BBN).
This concordance places an upper limit $\Omega_{GW}\lesssim10^{-5}$ on the total, i.e. 
integrated over all frequencies, energy of the gravitational wave background (see for example \cite{Brustein1996}). 
However, this limit assumes that the gravitational wave background was produced prior to the BBN. 
In contrast, the CMB experiments will also be sensitive to HFGW backgrounds produced at later epochs up to and around the period of recombination.
Moreover, CMB experiments {can} probe the gravitational wave background in a relatively narrow frequency bandwidth around $10^{10}~\rm{Hz}$
and are therefore sensitive to sharply peaked HFGW spectra whose total energy might not exceed the BBN limit. In addition, a dedicated CMB experiment
could improve sensitivity by 3-4 orders of magnitude, leading to a sensitivity comparable to the BBN limit. 
In any case, it is worth pointing out that CMB experiments provide an independent technique for observing or constraining HFGWs.
}


\section{Conclusion\label{SectionIV}}

In this work, we have analyzed the implications of graviton to photon conversion in the presence of large scale magnetic fields.
We have evaluated the conversion probability in the magnetic fields associated with galaxy clusters and filaments as well as primordial magnetic fields. Our estimation imply that this conversion probability is highest for primordial magnetic fields (assuming that PMF have a characteristic strength $B_0\sim10^{-9}~{\rm G}$). Assuming realistic values for the magnetic fields, we have shown that a sufficiently strong HFGW background would lead to an observable signature in frequency spectrum of the CMB. We argue that, this signature could be separated from other sources of variations in CMB like the SZ and galactic foregrounds using their corresponding frequency dependences. The current day CMB experiments allow to place significant constraints on the HFGW background ($\Omega_{GW}\lesssim1$). These limits are about 25 orders of magnitude stronger than {existing direct constraints in the high frequency region}. Furthermore, these limits could be improved by about 3-4 orders of magnitude in an experiment dedicated to constraining HFGWs.



\section*{Acknowledgements}
The authors thank Phil Mauskopf for useful discussions. This
research has made use of NASA's Astrophysics Data System.



\end{document}